\documentclass[pre,twocolumn,showpacs,amsmath,amssymb]{revtex4-1}
\usepackage{color}
\newcommand{\gae}{\lower 2pt \hbox{$\, \buildrel {\scriptstyle >}\over {\scriptstyle
\sim}\,$}}
\begin{document}
\def\be{\begin{equation}}
\def\ee{\end{equation}}
\def\bea{\begin{eqnarray}}
\def\eea{\end{eqnarray}}
\def\l{\label}
\def\fr{\frac}
\def\o{\omega}
\def\n{\nabla}
\def\a{\alpha}
\def\b{\beta}
\title {Quantum Langevin equation of a charged oscillator in a magnetic field and coupled to a heat bath through momentum variables}
\author{Shamik Gupta$^1$ and Malay Bandyopadhyay$^2$}
\affiliation{$^1$Laboratoire de Physique de l'\'{E}cole Normale Sup\'{e}rieure de Lyon, Universit\'{e}
de Lyon, CNRS, 46 All\'{e}e d'Italie, 69364 Lyon c\'{e}dex 07, France\\
$^2$Chemical Physics Theory Group, Department of Chemistry, University of Toronto, 80 St. George Street, Toronto, Ontario M5S 3H6, Canada}
\date{\today}
\begin{abstract}
We obtain the quantum Langevin equation (QLE) of a charged quantum
particle moving in a harmonic potential in the presence of a uniform external magnetic field and linearly coupled to a quantum heat bath through momentum variables. The bath is modeled as a collection of independent quantum harmonic oscillators. The QLE involves a random force which does not depend on the magnetic field, and a quantum-generalized classical Lorentz force. These features are also present in the QLE for the case of particle-bath coupling through coordinate variables. However, significant differences are also observed. For example, the mean force in the QLE is characterized by a memory function that depends explicitly on the magnetic field. The random force has a modified form with correlation and commutator different from those in the case of coordinate-coordinate coupling. Moreover, the coupling constants, in addition to appearing in the random force and in the mean force, also renormalize the inertial term and the harmonic potential term in the QLE. 
\end{abstract}
\maketitle
\section{Introduction}
The issue of the magnetic response of a charged quantum particle moving
in a potential arises in many problems of theoretical and experimental
relevance, e.g., Landau diamagnetism \cite{Landau:1930,vanVleck:1932},
quantum Hall effect \cite{Klitzing:1980,Laughlin:1981}, two-dimensional
electronic systems \cite{Jacak:1998}, and others. The additional effect
of quantum dissipation due to interaction with the external environment
may be studied systematically by employing the system-plus-reservoir
model, i.e., the Caldeira-Leggett ͑model \cite{Caldeira:1983} ͑(also
known as the independent-oscillator ͑model
\cite{Hakim:1985,Magalinskii:1959,Ford:1987,Ford:1988,Grabert:1987,Hanggi:2005}). In this scheme, the environment is modelled as a quantum mechanical heat bath or reservoir comprising an infinite number of independent quantum harmonic oscillators with continuously distributed frequencies. One assumes a specific coupling of the dynamical variables of the oscillators to those of the particle.  

In the case of bilinear coupling between the particle coordinate and the coordinate of each bath oscillator, a reduced description of the particle motion is given by the quantum Langevin equation (QLE) satisfied by the particle coordinate operator. In this equation, coupling to the bath is described by (i) an operator-valued random force, and (ii) a mean force characterized by a memory function \cite{Ford:1988, Ford:1990}. These forces do not depend on the magnetic field whose only appearance in the QLE is through a quantum generalization of the classical Lorentz force.
    
In this work, we consider the complementary possibility of coupling of a
quantum system to a quantum mechanical heat bath through the momentum
variables. Although such a scenario has been considered previously by
many authors \cite{Caldeira:1983, Leggett:1984, Cuccoli:2001, Bao:2005,
Bai:2005, Bao:2006, Ankerhold:2007}, here we study the additional
feature of the presence of an external magnetic field. To this end, we
consider a gauge-invariant system-plus-reservoir model. The system
comprises a charged quantum particle moving in a harmonic potential in
the presence of a magnetic field. The particle is linearly coupled via the momentum variables to a quantum heat bath consisting of independent quantum harmonic oscillators. 

Here, we derive a QLE for the particle coordinate operator for the case
of an external magnetic field which is uniform in space. The QLE is
obtained by utilizing the well-known Heisenberg equation of motion for
evolution of quantum operators and by effectively integrating out the
bath degrees of freedom from the equations of motion. We show that
similar to the case of coordinate-coordinate coupling, the QLE involves
(i) a quantum-generalized Lorentz force term, and (ii) a random force
which does not depend on the magnetic field. This latter force,
nevertheless, has a modified form, with symmetric correlation and
unequal time commutator different from the corresponding results in the
case of coordinate-coordinate coupling. Other differences include (i)
the memory function characterizing the mean force in the QLE has an
explicit dependence on the magnetic field, and (ii) the inertial term
and the harmonic potential term in the QLE get renormalized by the
coupling constants. 

The paper is organized as follows. In the next section, we introduce the system of study, and show that the system is invariant under a gauge transformation. In Section \ref{EOM}, we derive the Heisenberg equations of motion for the particle and the bath oscillators. In Section \ref{Special}, we derive the QLE for the charged particle for the case of a magnetic field which is uniform in space. Finally, we draw our conclusions.

\section{System of study}
\label{System}
Consider a charged particle moving in a harmonic potential in the presence of an external magnetic field. The particle is linearly coupled through the momentum variables to a large number $N$ of independent quantum harmonic oscillators constituting a heat bath. The Hamiltonian of the system is given by 
\bea
H&=&\fr{1}{2m}\Big({\bf p}-\fr{e}{c}{\bf A}\Big)^2+\fr{1}{2}m\o_0^2{\bf r}^2\nonumber \\
&&+\sum_{j=1}^N\Big[\fr{1}{2m_j}\Big({\bf p}_j-g_j{\bf
p}+\fr{g_je}{c}{\bf A}\Big)^2+\fr{1}{2}m_j\o_j^2{\bf q}_j^2\Big], \nonumber \\
\l{H}
\eea
where
$e,m,{\bf p},{\bf r}$ are respectively the charge, the mass, the
momentum operator and the coordinate operator of the particle, while
$\o_0$ is the frequency characterizing its motion in the harmonic
potential. The $j$th heat-bath oscillator has mass $m_j$, frequency
$\o_j$, coordinate operator ${\bf q}_j$, and momentum operator ${\bf
p}_j$. The dimensionless parameter $g_j$ describes the coupling between the particle and the $j$th oscillator. The speed of light in vacuum is denoted by $c$. The vector potential ${\bf A}={\bf A}({\bf r})$ is related to the external magnetic field ${\bf B}({\bf r})$ through
\be
{\bf B}({\bf r})=\n \times {\bf A}({\bf r}).
\l{B}
\ee

The relevant commutation relations for the various coordinate and momentum operators are
\be
[r_\a,p_\b]=i\hbar \delta_{\a \b}, ~~~~[q_{j\a},p_{k\b}]=i\hbar \delta_{jk}\delta_{\a \b},
\l{canonical-commutation}
\ee
while all other commutators vanish. In the above equation, $\delta_{jk}$
denotes the Kronecker $\delta$ function. Here, and in the following, Greek indices ($\a, \b, \ldots$) refer to the three spatial directions, while Roman indices ($i,j,k,\ldots$) represent the heat-bath oscillators.

We now show that our system of study is gauge invariant. Consider the gauge
transformation
\be
{\bf A}({\bf r}) \rightarrow {\bf A}'({\bf r})={\bf A}({\bf r})+\n f({\bf r}),
\l{gauge-transformation}
\ee
where $f({\bf r})$ is an arbitrary function of coordinate ${\bf r}$. The transformed Hamiltonian $H'=H'({\bf A}')$ is given by the right hand side of Eq. (\ref{H}) with ${\bf A}$ replaced by ${\bf A}'$. 

Now, our system will be gauge invariant if simultaneous with the transformation (\ref{gauge-transformation}), one can make a unitary transformation of the state vectors of the system,
\be
|\psi(t)\rangle \rightarrow |\psi'(t)\rangle=U|\psi(t)\rangle; ~~~~U^\dagger=U^{-1},
\ee 
such that all physical observables remain invariant under the joint transformation. This requires that one should have $H'({\bf A'})=UH({\bf A})U^\dagger$, where $H({\bf A})\equiv H$. In our case, finding such a unitary transformation is easily achieved with the choice
\be
U=\exp\Big(\fr{ie}{\hbar c}f({\bf r})\Big).
\ee
Using the Hadamard formula 
\be
e^X Y e^{-X}=Y+[X,Y]+\fr{1}{2!}[X,[X,Y]]+\ldots,
\ee
and the commutation relations (\ref{canonical-commutation}), one can check that $H'({\bf A}')=UH({\bf A})U^\dagger$, as required.
\section{Heisenberg equations of motion}
\label{EOM}
In this section, we derive the Heisenberg equations of motion for the
charged particle and the heat-bath oscillators.
\subsubsection{Charged particle}
For the charged particle, the Heisenberg equations of motion are
\bea
{\bf v} \equiv \dot{\bf r}&=&\fr{1}{i\hbar}[{\bf r},H]\nonumber \\
&=&\fr{1}{m}\Big({\bf p}-\fr{e}{c}{\bf A}\Big)-\sum_{j=1}^N
\fr{g_j}{m_j}\Big({\bf p}_j-g_j{\bf p}+\fr{g_je}{c}{\bf A}\Big),\nonumber \\
\l{rdot}
\eea
and
\bea
\dot p_\a&=&\fr{1}{i\hbar}[p_\a,H] \nonumber \\
&=&\fr{e}{2c}\Big[(\partial_\a A_\b)v_\b+v_\b(\partial_\a
A_\b)\Big]-m\o_0^2r_\a.
\l{palphamid1}
\eea
Equation (\ref{rdot}) gives
\be
\dot{\bf p}=m_{\rm r}\ddot{\bf r}+\fr{e}{c}\dot{\bf A}+\sum_{j=1}^N
\fr{g_jm_{\rm r}}{m_j}\dot{\bf
p}_j,
\l{pdot}
\ee
where $m_{\rm r}$ is the ``renormalized mass", defined as
\be
m_{\rm r}\equiv m/\Big[1+\sum_{j=1}^N \fr{g_j^2m}{m_j}\Big]. 
\l{mrdefinition}
\ee

Next, note that
\be
({\bf v} \times {\bf B})_\a=v_\b\partial_\a A_\b-v_\b\partial_\b A_\a,
\l{vcrossB}
\ee
and that
\bea
(\partial_\a A_\b)v_\b&=&v_\b(\partial_\a A_\b)+[\partial_\a A_\b,v_\b]
\nonumber \\
&=&v_\b(\partial_\a A_\b)+\fr{i\hbar}{m_{\rm r}} \partial_\a \partial_\b
A_\b.
\l{palphamid3}
\eea
Using Eqs. (\ref{vcrossB}) and (\ref{palphamid3}) in Eq.
(\ref{palphamid1}), we get
\bea
\dot p_\a&=&\fr{e}{c}({\bf v}\times {\bf B})_\a+\fr{e}{c}v_\b\partial_\b A_\a+\fr{i\hbar
e}{2m_{\rm r}c} \partial_\a \partial_\b A_\b-m\o_0^2r_\a, \nonumber \\
\eea
that is,
\bea
\dot{\bf p}&=&\fr{e}{c}({\bf v}\times {\bf B})+\fr{e}{c}({\bf v}.\n){\bf
A}+\fr{i\hbar e}{2m_{\rm r}c}\n (\n.{\bf A})-m\o_0^2 {\bf r}. \nonumber
\\
\l{pdotagain}
\eea

Now, we have
\bea
\dot A_\a&=&\fr{1}{i\hbar}[A_\a,H]\nonumber \\
&=&v_\b(\partial_\b
A_\a)+\fr{i\hbar}{2m_{\rm r}}\partial_\b\partial_\b
A_\a,
\l{Amid1}
\eea
so that
\be
\dot{\bf A}({\bf r})=({\bf v}.\n){\bf
A}+\fr{i\hbar}{2m_{\rm r}}\n^2
{\bf A},
\l{Adot}
\ee
which, on substituting in Eq. (\ref{pdot}), gives
\bea
\dot{\bf p}&=&m_{\rm r}\ddot{\bf r}+\fr{e}{c}({\bf v}.\n){\bf A}+\fr{i\hbar
e}{2m_{\rm r}c}\n^2 {\bf
A}+\sum_{j=1}^N \fr{g_jm_{\rm r}}{m_j}\dot{\bf p}_j. \nonumber \\
\l{pdotwithAdot}
\eea

On equating Eq. (\ref{pdotagain}) with Eq. (\ref{pdotwithAdot}), we get
\bea
m_{\rm r}\ddot {\bf r}&=&-m\o_0^2{\bf r}+\fr{e}{c}({\bf v}\times {\bf B})\nonumber \\
&&+\fr{i\hbar
e}{2m_{\rm r}c}\Big(\n(\n.{\bf A})-\n^2 {\bf A}\Big)-\sum_{j=1}^N
\fr{g_jm_{\rm r}}{m_j}\dot {\bf p}_j. \nonumber \\
\eea
On noting that
\be
\n(\n.{\bf A})-\n^2 {\bf A}=\n \times (\n \times {\bf A})=\n \times {\bf B}=\fr{4\pi}{c}{\bf j},
\l{jreason}
\ee
where ${\bf j}$ is the current producing the external magnetic field,
and also the fact that in practice this current source lies outside the region where the charged particle moves, we have 
\bea
m_{\rm r}\ddot {\bf r}&=&-m\o_0^2{\bf r}+\fr{e}{c}({\bf v}\times {\bf
B})-\sum_{j=1}^N \fr{g_jm_{\rm r}}{m_j}\dot {\bf p}_j.
\eea
In the next subsection, we show that $\dot{\bf p}_j=-m_j\o_j^2{\bf q}_j$. Using this in the last equation, we get
\be
m_{\rm r}\ddot {\bf r}=-m\o_0^2{\bf r}+\fr{e}{c}({\bf v}\times {\bf
B})+\sum_{j=1}^N g_jm_{\rm r}\o_j^2{\bf q}_j.
\l{heqmsystem}
\ee
\subsubsection{Heat-bath oscillators}
For the heat-bath oscillators, the equations of motion are
\bea
\dot{\bf q}_j&=&\fr{1}{i\hbar}[{\bf q}_j,H]\nonumber \\
&=&\fr{1}{m_j}\Big({\bf p}_j-g_j{\bf p}+\fr{g_je}{c}{\bf
A}\Big),
\l{qjdot}
\eea
and
\bea
\dot{\bf p}_j&=&\fr{1}{i\hbar}[{\bf p}_j,H]\nonumber\\
&=&-m_j\o_j^2{\bf q}_j.
\l{pjdot}
\eea

Combining Eqs. (\ref{qjdot}) and (\ref{pjdot}), we get
\be
m_j\ddot{\bf q}_j=-m_j\o_j^2{\bf q}_j-g_j\dot{\bf
p}+\fr{g_je}{c}\dot {\bf A},
\l{qjddot}
\ee
which, on using Eqs. (\ref{pdotagain}) and (\ref{Adot}), gives
\bea
m_j\ddot{\bf q}_j&=&-m_j\o_j^2{\bf q}_j+g_jm\o_0^2{\bf r}-\fr{g_j e}{c}({\bf v} \times {\bf B})\nonumber \\
&&+\fr{i\hbar
g_je}{2m_{\rm r}c}\Big(\n^2
{\bf A}-\n (\n.{\bf A})\Big). \nonumber \\
\eea
Using Eq. (\ref{jreason}) and the reasoning given in the sentence following it, we finally have
\be
m_j\ddot{\bf q}_j=-m_j\o_j^2{\bf q}_j+g_jm\o_0^2{\bf r}-\fr{g_j e}{c}({\bf v} \times {\bf B}).
\l{heat-bath-EOM-Bz}
\ee
\section{Uniform ${\bf B}$: The quantum Langevin equation}
\label{Special}
In this section, we derive a QLE for the charged particle interacting
with the heat-bath oscillators as modelled by Eq. (\ref{H}), where we
now consider a magnetic field uniform in space. One of the early
appearances of a QLE in the case of coordinate-coordinate coupling between
the particle and the heat-bath oscillators in the absence of magnetic
field was in Ref.
\cite{Magalinskii:1959}. In our case, we follow the program adopted in
\cite{Ford:1990} for the derivation of the QLE. The essential steps are
as follows.
\begin{itemize}
\item{Step $1$: Obtain the Heisenberg equations of motion for the system of the charged particle coupled to the heat bath. Solve these equations for the bath variables, and substitute the solution into the equations for the charged particle to obtain a reduced description of the particle motion. The solution will contain explicit expressions for the dynamical variables at time $t$ in terms of their initial values.}
\item{Step $2$: Make specific assumptions about the initial state of the system, e.g., assume that the heat bath was at thermal equilibrium at the initial instant with the bath variables distributed according to a canonical distribution.}
\item{Step $3$: Show that the coordinate operator for the charged particle then represents a stochastic process in time, and satisfies a QLE. The statistical properties of the stochastic process arise from the initial canonical distribution of the heat bath.}
\end{itemize}

Step $1$ has been partially carried out in Sec. \ref{EOM}. We now
carry out the remaining part, and solve the equations of motion for the
bath variables by considering the magnetic field ${\bf B}$ to be uniform
in space, with components $B_x, B_y, B_z$, and magnitude $B=\sqrt{B_x^2+B_y^2+B_z^2}$.
In this case, Eq. (\ref{heat-bath-EOM-Bz}) has the retarded solution
\bea
{\bf q}_j(t)&=&{\bf q}^h_j(t)+\fr{g_jm\o_0^2}{m_j\o_j^2}{\bf
r}(t)-\fr{g_jm\o_0^2}{m_j\o_j^2}{\bf r}(0)\cos(\o_j t)\nonumber\\
&&- \fr{g_jm\o_0^2}{m_j\o_j^2}\int_0^t dt'~\dot{{\bf r}}(t')\cos(\o_j(t-t'))\nonumber\\
&&-\fr{g_jm\o_c}{m_j\o_j B}\Gamma \int_0^t dt'~\dot{{\bf r}}(t')\sin(\o_j(t-t')),
\l{heat-bath-solution}
\eea
where 
\be
{\bf q}^h_j(t)\equiv{\bf q}_j(0)\cos(\o_jt)+\fr{{\bf
p}_j(0)}{m_j\o_j}\sin(\o_jt)
\ee
is the contribution from the initial condition,
\be
\o_c\equiv\fr{e B}{mc}
\ee
is the Larmor frequency of precessional motion of the charged particle in the magnetic field, and
\be
\Gamma\equiv\begin{bmatrix}
  0 & B_z & -B_y \\
  -B_z & 0 & B_x \\
  B_y & -B_x & 0
 \end{bmatrix}. \\
\ee

Substituting Eq. (\ref{heat-bath-solution}) into Eq. (\ref{heqmsystem}), we
get 
\bea
&&m_{\rm r}\ddot {\bf r}+\int_0^t dt'\dot{\bf r}(t')\mu(t-t')+m_{\rm
r}\o_0^2{\bf r}+\mu_{\rm d}(t){\bf r}(0)\nonumber \\
&&-\fr{e}{c}({\bf v}\times {\bf B})={\bf F}(t),
\l{eqnofmotion}
\eea
where
\bea
{\bf F}(t)&\equiv&\sum_{j=1}^N g_jm_{\rm r}\o_j^2{\bf
q}^h_j(t)\Theta(t), \l{Ft} \\
\mu(t-t')&\equiv&\mu_{\rm d}(t-t')+\Gamma \mu_{\rm od}(t-t'),
\eea
where $\mu_{\rm d}$, the diagonal part of function $\mu$, and $\mu_{\rm od}$, the off-diagonal part, are given by
\bea
&&\mu_{\rm d}(t-t')\equiv\sum_{j=1}^N \fr{g_j^2 m m_{\rm r}\o_0^2}{m_j}\cos(\o_j(t-t'))\Theta(t-t'),\nonumber \\ \l{mu-d} \\
&&\mu_{\rm od}(t-t')\equiv\sum_{j=1}^N \fr{g_j^2 m m_{\rm
r}\o_j\o_c}{m_jB}\sin(\o_j(t-t'))\Theta(t-t'). \nonumber \\ \l{mu-od}   
\eea
This completes step $1$ of the program.

To implement step $2$, we now assume that at distant past, $t=-\infty$,
there was no magnetic field, the charged particle was held fixed at
${\bf r}(0)$, while the heat-bath oscillators were kept in weak contact
with another heat bath at temperature $T$ so as to be able to come to
thermal equilibrium. Therefore, at time $t=0$, the heat-bath oscillators
are in canonical equilibrium at temperature $T$ with respect to the free oscillator Hamiltonian
\bea
H_{\rm B}=\sum_{j=1}^N\Big[\fr{{\bf p}^2_j}{2m_j}+\fr{1}{2}m_j\o_j^2{\bf q}_j^2\Big].
\l{HB}
\eea
Subsequently, at a time $t \gae 0$, the particle is released and the
magnetic field is turned on, so that further evolution of the system is
governed by Hamiltonian (\ref{H}). Note that this physical picture
is consistent with choosing the retarded solution
(\ref{heat-bath-solution}). The state of the system at
$t=0$, corresponding to a correlation-free preparation, is given by the total density matrix operator 
\be
\rho_{\rm T}(0)=\rho_{\rm P}(0)\bigotimes\rho_{\rm B},
\l{rhoT-initial}
\ee
where the initial density matrix operator $\rho_{\rm P}(0)$ of the charged particle is given by
\be
\rho_{\rm P}(0)=\delta\Big({\bf r}-{\bf r}(0)\Big)\delta({\bf p}),
\l{rho-particle-initial}
\ee
while that of the heat bath, which is in canonical equilibrium, is given by  
\be
\rho_{\rm B}=\fr{e^{-H_{\rm B}/k_BT}}{Z_{\rm B}}; ~~~~Z_{\rm B}={\rm
Tr}_{\rm B}(e^{-H_{\rm B}/k_BT}).
\l{rho-bath}
\ee
Here, $k_B$ is the Boltzmann constant. The normalization factor is
denoted by $Z_{\rm B}$, while ${\rm Tr}_{\rm B}$ represents partial trace
operation with respect to the bath variables.

The statistical average of a heat-bath operator $O$ with respect to the
initial state (\ref{rhoT-initial}) is given by
\be
\langle O \rangle \equiv {\rm Tr}_{\rm B}(Oe^{-H_{\rm B}/k_BT})/{\rm
Tr}_{\rm B}(e^{-H_{\rm B}/k_BT}).
\ee
Using known properties of quantum harmonic oscillators, it is straightforward to show that
\bea
&&\langle q_{j\a}(0) \rangle=0, \nonumber \\
&&\langle p_{j\a}(0) \rangle=0, \nonumber \\
&&\langle q_{j\a}(0) q_{k\b}(0) \rangle=\frac{\hbar}{2m_j\o_j}\coth\Big(\frac{\hbar\o_j}{2k_BT}\Big)\delta_{jk}\delta_{\a\b}, \nonumber \\ 
&&\langle p_{j\a}(0) p_{k\b}(0) \rangle=\frac{\hbar m_j\o_j}{2}\coth\Big(\frac{\hbar\o_j}{2k_BT}\Big)\delta_{jk}\delta_{\a\b}, \nonumber \\
&&\langle q_{j\a}(0) p_{k\b}(0) \rangle=-\langle p_{j\a}(0) q_{k\b}(0) \rangle=\fr{1}{2}i\hbar\delta_{jk}\delta_{\a\b}. \nonumber \\
\l{bath-canonical-average}
\eea
In addition, we have the Gaussian property: the statistical average of
an odd number of factors of $q_{j\a}(0)$ and $p_{j\a}(0)$ is zero, while
that of an even number of factors is equal to the sum of products of
pair averages with the order of the factors preserved. 

Using the results in Eq. (\ref{bath-canonical-average}), one finds that the force operator ${\bf F}(t)$, Eq. (\ref{Ft}), has zero mean,
\be
\langle {\bf F}(t) \rangle=0,
\ee
and a symmetric correlation given by
\bea
&&\frac{1}{2}\langle F_\a(t) F_\b(t')+ F_\b(t') F_\a(t) \rangle \nonumber \\
&&=\frac{\hbar\delta_{\a,\b}}{2}\sum_{j=1}^N\fr{g_j^2 m_{\rm r}^2\o_j^3}{m_j}\coth\Big(\frac{\hbar\o_j}{2k_BT}\Big)\cos(\o_j(t-t')).\nonumber \\
\l{Ft-symmetric-correlation}
\eea
In addition, ${\bf F}(t)$ has the Gaussian property, which follows from
the same property of the ${\bf q}_j(0)$ and ${\bf p}_j(0)$.
 
Thus, the initial distribution of the heat bath oscillators turns the force operator ${\bf F}(t)$ into an operator-valued random force. 
On using the canonical commutation rules (\ref{canonical-commutation}), we find that ${\bf F}(t)$ has the unequal time commutator given by
\be
[F_\a(t),F_\b(t')] = -i\hbar\delta_{\a,\b}\sum_{j=1}^N \fr{g_j^2 m_{\rm r}^2 \o_j^3}{m_j}\sin(\o_j(t-t')).
\l{Ft-unequal-time-commutator}
\ee

We are now in a position to achieve Step $3$ and interpret Eq. (\ref{eqnofmotion}) with $t \gae 0$ as a QLE for the particle coordinate operator ${\bf r}(t)$, which now reads 
\bea
&&m_{\rm r}\ddot {\bf r}+\int_0^t dt'\dot{\bf r}(t')\mu(t-t')+m_{\rm
r}\o_0^2{\bf r}+\mu_{\rm d}(t){\bf r}(0)\nonumber \\
&&-\fr{e}{c}({\bf v}\times {\bf B})={\bf F}(t),
\l{QLE}
\eea
where ${\bf F}(t)$ represents a random force with correlation and
unequal time commutator given by Eqs. (\ref{Ft-symmetric-correlation})
and (\ref{Ft-unequal-time-commutator}), respectively. The renormalized
mass $m_{\rm r}$ is given by Eq. (\ref{mrdefinition}). The second term
on the left represents a mean force characterized by the friction kernel
or the memory function $\mu(t)$. Note the appearance of the initial
value term that depends explicitly on the
initial coordinate of the particle and the diagonal part of the 
memory function. One can absorb this term into the
definition of the random force by defining ${\bf G}(t)={\bf F}(t)-\mu_{\rm
d}(t){\bf r}(0)$, and then considering the initial state (\ref{rhoT-initial}),
with particle density operator (\ref{rho-particle-initial}) and bath
density operator ${\rho}_{\rm B}=\fr{e^{-H^{\rm Shifted}_{\rm
B}/k_BT}}{Z_{\rm B}}$, where the ``shifted" bath Hamiltonian is
$H^{\rm Shifted}_{\rm B}=\sum_{j=1}^N \Big[\frac{{\bf p}_j^2}{2m_j}+\frac{1}{2} 
m_j\omega_j^2 [{\bf q}_j-\frac{g_j m\omega_0^2}{m_j\omega_j^2}{\bf 
r}(0)]^2\Big]$ \cite{Hanggi:2005}. This procedure guarantees that the
redefined random force ${\bf G}(t)$ has the same statistical properties as ${\bf F}(t)$.

We now point out some interesting features of the QLE (\ref{QLE}), which are not present in the QLE for the case of coordinate-coordinate coupling \cite{Ford:1990}. These are (i) The coupling renormalizes the inertial mass, (ii) the harmonic potential term is also renormalized, (iii) the friction kernel has an off-diagonal part arising from the magnetic field and a diagonal part due to the harmonic potential. Similar to the coordinate-coordinate coupling, the magnetic field appears in the QLE as a quantum-generalized classical Lorentz force term, and the random force in the QLE does not depend on the magnetic field. This latter force, nevertheless, has a different form so that its symmetric correlation and unequal time commutator are modified from the corresponding expressions in the case of coordinate-coordinate coupling.

It is interesting to see that the correlation and commutator of the random force ${\bf F}(t)$ may be related to the friction kernel $\mu(t-t')$. The Laplace transform of its diagonal part $\mu_{\rm d}(t)$, Eq. (\ref{mu-d}), is given by
\bea
\widetilde{\mu}_{\rm d}(\o)&\equiv&\int_0^{\infty} dt ~\mu(t)e^{i\o t}; ~~~~{\rm Im}(\o)>0\nonumber \\
&=&\sum_{j=1}^N\frac{g_j^2m m_{\rm r}\o_0^2}{m_j}\int_0^{\infty} dt \cos(\o_jt)e^{i\o t} \nonumber \\
&=&\fr{i}{2}\sum_{j=1}^N\fr{g_j^2m m_{\rm r}\o_0^2}{m_j}\Big(\frac{1}{\o-\o_j}+\frac{1}{\o+\o_j}\Big).
\eea
Using the well-known result that $1/(x+i0^+)=P(1/x)-i\pi\delta(x)$, we have
\bea
{\rm Re}[\widetilde{\mu}_{\rm
d}(\o+i0^+)]&=&\fr{\pi}{2}\sum_{j=1}^N\fr{g_j^2m m_{\rm r}\o_0^2}{m_j}\nonumber \\
&&~~~~\times\Big(\delta(\o-\o_j)+\delta(\o+\o_j)\Big),\nonumber \\
\eea
so that Eq. (\ref{Ft-symmetric-correlation}) may be rewritten as
\bea
&&\fr{1}{2}\langle F_\a(t) F_\b(t')+F_\b(t')F_\a(t) \rangle \nonumber \\
&&=\fr{\hbar\delta_{\a,\b}}{\pi}\int_0^{\infty}d\o~{\rm
Re}[\widetilde{\mu}_{\rm d}(\o+i0^+)]\fr{\o^3m_{\rm r}}{\o_0^2m}\nonumber \\
&&~~~~~~~~\times\coth\Big(\fr{\hbar\o}{2k_BT}\Big)\cos(\o(t-t')),
\eea
and similarly, Eq. (\ref{Ft-unequal-time-commutator}) as
\bea
[F_\a(t),F_\b(t')]&=&\fr{2\hbar \delta_{\a,\b}}{i\pi}\int_0^{\infty}d\o ~{\rm
Re}[\widetilde{\mu}_{\rm d}(\o+i0^+)]\fr{\o^3m_{\rm r}}{\o_0^2m}\nonumber \\
&&~~~~~~~~\times\sin(\o(t-t')).
\eea

\section{Conclusions}
\label{Conclusions}
In this work, we derived a quantum Langevin equation (QLE) for a charged
quantum particle moving in a harmonic potential in the presence of a uniform external magnetic field and coupled linearly through the momentum variables to a collection of independent quantum harmonic oscillators constituting a heat bath. In this QLE, the magnetic field appears through a quantum-generalized classical Lorentz force term. The QLE involves a random force which does not depend on the magnetic field. These aspects are also present in the QLE for the case of particle-bath coordinate-coordinate coupling \cite{Ford:1990}. However, significant differences are also observed: (i) The random force has a modified form with symmetric correlation and unequal time commutator different from those in the case of coordinate-coordinate coupling, (ii) the inertial term and the harmonic potential term in the QLE get renormalized, and (iii) the memory function characterizing the mean force in the QLE has a field-independent diagonal part, but also an explicit field-dependent off-diagonal part. 
\begin{acknowledgments}
S. G. thanks the Weizmann Institute of
Science, Israel, where this work was initiated. He acknowledges support of the Israel Science
Foundation and the French Contract No. ANR-10-CEXC-010-01. M. B. acknowledges
support of the Connaught fund and the NSERC. 
\end{acknowledgments}


\begin{thebibliography}{99}
\bibitem{Landau:1930}L. Landau, Z. Phys. {\bf 64}, 629 (1930).
\bibitem{vanVleck:1932}J. H. Van Vleck, {\em The Theory of Electric and Magnetic Susceptibilities} (Oxford University Press, London, 1932).
\bibitem{Klitzing:1980}K. v. Klitzing, G. Dorda, and M. Pepper, Phys.
Rev. Lett. {\bf 45}, 494 (1980).
\bibitem{Laughlin:1981}R. B. Laughlin, Phys. Rev. B {\bf 23}, 5632 (1981).
\bibitem{Jacak:1998}L. Jacak, P. Hawrylak, and A. W\'{o}js, {\em Quantum
dots} (Springer-Verlag, Berlin, 1998).
\bibitem{Caldeira:1983}A. O. Caldeira and A. J. Leggett, Physica A {\bf 121}, 587 (1983); Ann. Phys. {\bf 149}, 374 ͑(1983͒).
\bibitem{Magalinskii:1959}V. B. Magalinskii, Zh. Eksp. Teor. Fiz. {\bf
36}, 1942 (1959) [Sov. Phys. JETP {\bf 9}, 1381 (1959)].
\bibitem{Hakim:1985}V. Hakim and V. Ambegaokar, Phys. Rev. A {\bf 32}, 423 (1985).
\bibitem{Ford:1987}G. W. Ford and M. Kac, J. Stat. Phys. {\bf 46}, 803 (1987).
\bibitem{Grabert:1987}H. Grabert, P. Schramm, and G.-L. Ingold, Phys. Rev. Lett. {\bf 58}, 1285 (1987).
\bibitem{Ford:1988}G. W. Ford, J. T. Lewis, and R. F. O'Connell, Phys. Rev. A {\bf 37}, 4419 (1988).
\bibitem{Hanggi:2005}P. Hanggi and G.-L. Ingold, Chaos {\bf 15}, 026105
(2005).
\bibitem{Ford:1990}X. L. Li, G. W. Ford, and R. F. O'Connell, Phys. Rev. A {\bf 41}, 5287 (1990); Phys. Rev. A {\bf 42}, 4519 (1990).
\bibitem{Leggett:1984}A. J. Leggett, Phys. Rev. B {\bf 30}, 1208 (1984).
\bibitem{Cuccoli:2001}A. Cuccoli, A. Fubini, V. Tognetti, and R. Vaia, Phys. Rev. E {\bf 64}, 066124 (2001).
\bibitem{Bao:2005}J. D. Bao and Y. Z. Zhuo, Phys. Rev. E {\bf 71}, 010102(R) (2005).
\bibitem{Bai:2005}Z. W. Bai, J. D. Bao, and Y. L. Song, Phys. Rev. E {\bf 72}, 061105 (2005).
\bibitem{Bao:2006}J. D. Bao, Y. Z. Zhuo, F. A. Oliveira, and P. H\"{a}nggi, Phys. Rev. E {\bf 74}, 061111 (2006).
\bibitem{Ankerhold:2007}J. Ankerhold and E Pollak, Phys. Rev. E {\bf 75}, 041103 (2007).
\end{thebibliography}
\end{document}